\documentclass{article}
\usepackage{amsmath}
\usepackage{amssymb}

\usepackage{graphicx}

\begin{document}

\title{New Stabilization of the Burnett Equations when Entropy
Change to $K\negthinspace n^0$ Vanishes}
\author{Lars H. S\"{o}derholm \\
KTH, Mekanik, Stockholm}
\date{\today}
\maketitle

\begin{abstract}
We assume that to zero order in the Knudsen number the deviation of the 
entropy from a background value vansishes. We then show that adding a super-Burnett
term we obtain a stable state of rest. The resulting equations have the same
form as the Burnett equations but with the value of some
coefficients changed. In particular the result applies to nonlinear
acoustics.
\end{abstract}

We consider a slightly rarefied gas. To first order in the Knudsen number, $K \negthinspace n$, 
the Navier-Stokes equations are valid. Burnett \cite{Burnett} derived the corresponding 
equations to second order in $K \negthinspace n$. Bobylev \cite{Bobylev} 
showed that the state of rest is unstable for the
Burnett equations, see also Uribe et al. \cite{Uribe}. In this contribution
we make the assumption that the deviation of entropy from a background value
is of the order of $K \negthinspace n M \negthinspace a$, where $M \negthinspace a$, is the Mach number. This
is the case for nonlinear acoustics, where $M \negthinspace a\sim K \negthinspace n$. We show that with 
an error $K\negthinspace n^3$ the Burnett equations in this case can be replace by equations which are
linearly stable. 

In the one-dimensional case, the Burnett expressions for the $xx$ component
of the pressure tensor $P$ and the heat current $q$ are, see Chapman \&
Cowling \cite{CC} (the dots indicate nonlinear Burnett terms)\bigskip 
\begin{eqnarray*}
P &=&\frac{\rho T}{m}-\frac{4\mu }{3}v_{x}-\frac{2}{3}\frac{\mu ^{2}}{\rho }%
[\omega _{2}\frac{\rho _{xx}}{\rho }+(\omega _{2}-\omega _{3})\frac{T_{xx}}{T%
}]..., \\
q &=&-\kappa T_{x}-\frac{2}{3}\frac{\mu ^{2}}{\rho }(\theta _{2}-\theta
_{4})v_{xx}...
\end{eqnarray*}%
We now linearize around a state at rest and uniform temperature and density,
writing%
\begin{equation*}
T=T_{0}[1+\tilde{T}],\;\rho =\rho _{0}[1+\tilde{\rho}],\;v=\sqrt{\frac{%
k_{B}T_{0}}{m}}\tilde{v}.
\end{equation*}%
We introduce dimensionless variables, where the unit of length is of the
order of the mean free path

\begin{equation*}
x=x^{\ast }\frac{\mu _{0}}{\rho _{0}}\sqrt{\frac{m}{k_{B}T_{0}}},\;t=t^{\ast
}\frac{\mu _{0}}{\rho _{0}}\frac{m}{k_{B}T_{0}}.
\end{equation*}%
In the sequel stars and tildes are omitted. We obtain the linearized
one-dimensional Burnett equations 
\begin{eqnarray}
\rho _{t}+v_{x} &=&0,  \label{continuity} \\
v_{t} &=&-(\rho +T)_{x}+\frac{4}{3}v_{xx}+\frac{2}{3}\omega _{2}\rho _{xxx}-%
\frac{2}{3}(\omega _{3}-\omega _{2})T_{xxx},  \label{momentum} \\
\frac{3}{2}T_{t} &=&-v_{x}+\frac{3}{2}fT_{xx}-\frac{2}{3}(\theta _{4}-\theta
_{2})v_{xxx}.  \label{energy}
\end{eqnarray}%
$f=2m\kappa /3k_{B}\mu $ is the Eucken number. In the calculations we use
the value $f=5/2$. This is the lowest approximation in terms of Sonine
polynomial expansion for any interatomic potential and is experimentally
found to be a good approximation, see \cite{CC}.

Now we assume that the entropy change is of $0(K \negthinspace nM \negthinspace a).$ We then have 
\begin{equation*}
\frac{dT}{T}=(\gamma -1)\frac{d\rho }{\rho }+0(K \negthinspace nM \negthinspace a).
\end{equation*}%
($\gamma =c_{p}/c_{v}$). Linearizing and using dimensionless units we find 
\begin{equation*}
(\gamma -1)\rho _{xxx}-T_{xxx}=0(K \negthinspace nM \negthinspace a).
\end{equation*}%
Hence, to within terms $0(K \negthinspace n)$ we have for any value $\alpha $ which is $%
0(1) $%
\begin{equation*}
\omega _{2}\rho _{xxx}+(\omega _{2}-\omega _{3})T_{xxx}=[\omega _{2}+\alpha
(\gamma -1)]\rho _{xxx}+\frac{2}{3}(\omega _{2}-\omega _{3}-\alpha
)T_{xxx}+0(K \negthinspace nM \negthinspace a).
\end{equation*}%
Thus, we can change the values of $\omega _{2}$, $\omega _{3}$ to $\breve{%
\omega}_{2}$ och $\breve{\omega}_{3}$ in the linear part of the Burnett
contribution. 
\begin{eqnarray}
\breve{\omega}_{2} &=&\omega _{2}+\alpha (\gamma -1),
\label{nytt varde omega2} \\
\breve{\omega}_{3} &=&\omega _{3}+\alpha \gamma .  \label{nytt varde omega3}
\end{eqnarray}

Let us now choose $\alpha $ so that the coefficient of $\rho _{xxx}$
vanishes, or $\breve{\omega}_{2}=0,$ This gives for a monatomic gas 
\begin{equation*}
\breve{\omega}_{2}=0,\;\breve{\omega}_{3}=\omega _{3}-\frac{5}{2}\omega
_{2}..
\end{equation*}%
For Maxwell molcules 
\begin{equation*}
\omega _{2}=2,\omega _{3}=3;\;\breve{\omega}_{2}=0,\;\breve{\omega}_{3}=-2.
\end{equation*}%
For hard spheres%
\begin{equation*}
\omega _{2}=2.028,\omega _{3}=2.418,\;\breve{\omega}_{2}=0,\;\breve{\omega}%
_{3}=-2.652.
\end{equation*}%
As a consequence, the $\rho _{xxx}$ term disappears. The sign of the $%
T_{xxx} $ term changes.

For solutions proportional to $\exp [ikx+\Lambda t]$ we find

\begin{figure}
\includegraphics[width=8cm]{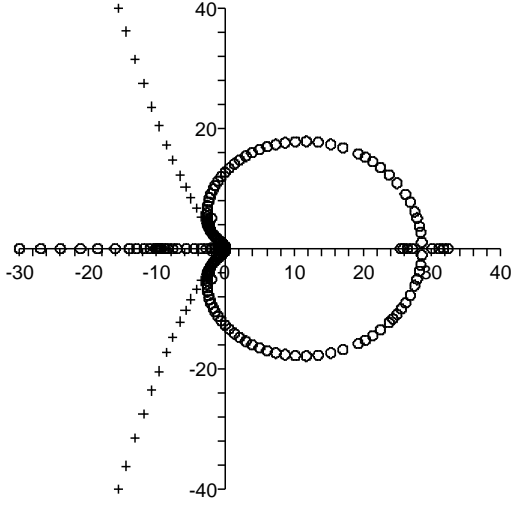}
\caption{Complex growth factor $\Lambda $
for hard spheres, $0<k<6$. Rings Burnett, crosses our equations}
\end{figure}

Asymptotically, for $k\rightarrow \infty $ we have, when $(\theta
_{2}-\theta _{4})(\check{\omega}_{2}-\check{\omega}_{3})>0$, 
\begin{eqnarray*}
\Lambda &=&-\frac{27}{8(\theta _{2}-\theta _{4})(\check{\omega}_{2}-\check{%
\omega}_{3})}(\frac{2}{3}\check{\omega}_{2}+\frac{1}{k^{2}})f, \\
\Lambda &=&\pm i\sqrt{\frac{8}{27}(\theta _{2}-\theta _{4})(\check{\omega}%
_{2}-\check{\omega}_{3})}k^{3}-\frac{(3f+4)}{6}k^{2}.
\end{eqnarray*}%
Clearly, there is one mode that is nonpropagating and damped and there are
two propagating, damped modes. One entropy mode and two sound wave modes. It
is really not necessary to have $\check{\omega}_{2}=0$, but just to have $%
\check{\omega}_{2}-\check{\omega}_{3}>0.$ 

Let us write down the resulting
equations, first in the one-dimensional case. We neglect the nonlinear Burnett terms. %
\begin{eqnarray*}
\rho _{t}+(\rho v)_{x} =0&& \\
\rho (v_{t}+vv_{x}) =-\frac{1}{m}(\rho T)_{x}+\frac{4}{3}(\mu v_{x})_{x}+%
\frac{2}{3}(\omega _{3}-\frac{5}{2}\omega _{2})\frac{\mu ^{2}}{\rho T}T_{xxx}&&
\\
\frac{3}{2m}\rho (T_{t}+vT_{x}) =-\frac{1}{m}\rho Tv_{x}+(\kappa
T_{x})_{x}+\frac{2}{3}(\theta _{4}-\theta _{2})\frac{\mu ^{2}}{\rho }v_{xxx}&&
\end{eqnarray*}%
Here, the coefficients of the Burnett terms can be taken at the background
value, but the variations of $\mu ,\kappa $ in the Navier-Stokes terms have
to be taken inte account. 

We now give the phase velocity. Note that the phase velocity
is constant plus a term to order $K\negthinspace n^2$. Hence the deviation 
from straight lines for the Navier-Stokes equations is not physically relevant
but that the deviation for the Burnett equations (and our equations) is. Note that
the difference between the Burnett equations and our equations is for
larger $k$ than those shown in Fig. 2.  
 
\begin{figure}
\includegraphics[width=8cm]{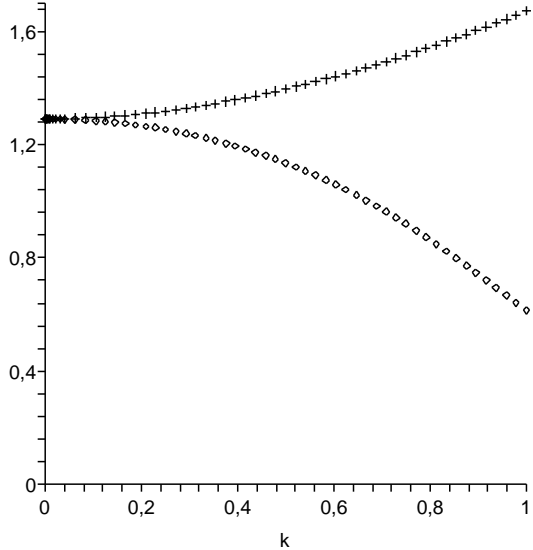}
\caption{Phase veloctiy for $0\leq
k\leq 1$. Hard spheres. Diamonds Navier-Stokes, crosses our equations}
\end{figure}

The three-dimensional equations are 
\begin{eqnarray*}
\rho _{t}+\mathbf{\nabla }\cdot (\rho v)_{x} =0,&& \\
\rho (\mathbf{v}_{t}+\mathbf{(v\cdot \nabla )v}) =-\frac{1}{m}\mathbf{%
\nabla }(\rho T),&& \\
+\mathbf{\nabla }\cdot \{\mu \lbrack \mathbf{\nabla v}+(\mathbf{\nabla v}%
)^{T}-\frac{2}{3}(\mathbf{\nabla \cdot v})1]\}+\frac{2}{3}(\omega _{3}-\frac{%
5}{2}\omega _{2})\frac{\mu ^{2}}{\rho T}\triangle (\mathbf{\nabla }T)&& \\
\frac{3}{2m}\rho (T_{t}+\mathbf{v\cdot \nabla }T) =-\frac{1}{m}\rho T(%
\mathbf{\nabla \cdot v})+\mathbf{\nabla \cdot }(\kappa \mathbf{\nabla }T)+%
\frac{2}{3}(\theta _{4}-\theta _{2})\frac{\mu ^{2}}{\rho }\triangle (\mathbf{%
\nabla \cdot v}).&&
\end{eqnarray*}

In earlier contributions by Jin and Slemrod, \cite{Jin} and by the present
author \cite{Soderholm},see also \cite{Stromgren} and \cite{Svard} the Burnett
equations were regularized to a set of 13 first order equations generally
valid. 

The present regularization applies when the deviations of entropy is $%
0(K \negthinspace nM \negthinspace a)$ but gives equations which can more easily be applied for small
Knudsen numbers. The condition on the entropy applies for nonlinear sound
propagation. The same assumption that $M \negthinspace a\sim K \negthinspace n$ is called the weakly
nonlinear case in Sone \cite{Sone}, where stationarity is assumed, but here
sound waves are included as well. 

Recently, the present author has also obtained another set of regularized 
equations \cite{Soderholm hybrid}, which like the Burnett equations and the equations in this work
are equations for $\rho, \mathbf{v}, T$. They are, however generally valid, with 
no limitation on entropy or Mach number. They are 
first order in time and third order in space, but also contain mixed derivatives
first order in time and up to second order in space.

\end{document}